# Pluto's Seasons: New Predictions for New Horizons

L. A. Young

Southwest Research Institute, Boulder, CO 80402

## ABSTRACT

Since the last Pluto volatile transport models were published (Hansen and Paige 1996), we have (i) new stellar occultation data from 2002 and 2006–2012 that have roughly twice the pressure as the discovery occultation of 1988, (ii) new information about the surface properties of Pluto, (iii) a spacecraft due to arrive at Pluto in 2015, and (iv) a new volatile transport model that is rapid enough to allow a large parameter-space search. Such a parameter-space search coarsely constrained by occultation results reveals three broad solutions: a high-thermal inertia, large volatile inventory solution with permanent northern volatiles (PNV); a lower thermal-inertia, smaller volatile inventory solution with exchanges between hemispheres, and a pressure plateau beyond 2015 (exchange with pressure plateau, EPP); and solutions with still smaller volatile inventories, with an early collapse of the atmosphere prior to 2015 (exchange with early collapse, EEC). PNV is favored by stellar occultation data, but EEC cannot yet be definitively ruled out without more atmospheric modeling or additional occultation observations and analysis.

Subject headings: Kuiper Belt — planets and satellites: individual (Pluto)

# 1. Introduction

Because Pluto's predominately $N_2$ atmosphere is in vapor-pressure equilibrium with the solid $N_2$ ice on its surface, the surface pressure is a sensitive function of the $N_2$ ice temperature. Furthermore, volatiles migrate from areas of higher insolation to areas of lower insolation, carrying both mass and latent heat (Stern et al., 1988, Spencer et al., 1997). The combination of Pluto's changing heliocentric distance and subsolar latitude leads to complex changes in Pluto's volatile distribution and surface pressure over its season. The first realistic models of Pluto's seasonal change were constructed in the mid 1990's (Hansen & Paige, 1996), post-dating the discovery of Pluto's atmosphere (Hubbard et al., 1989; Elliot et al., 1989), the identification of $N_2$ as the dominant volatile on the surface and in the atmosphere (Owen et al., 1993), and maps of the sub-Charon face of Pluto from mutual events (Buie et al., 1992; Young & Binzel, 1993). Most of the simulations in Hansen & Paige (1996) predicted large changes in Pluto's atmospheric pressure on decadal timescales.

New observational constraints postdating this model include occultations in 2002 and 2006–2012 (e.g., Elliot et al., 2003; Sicardy et al., 2003; Young et al., 2008; see Table 2), global albedo maps from HST observations in 1994 and 2002-2003 (Stern et al., 1997; Buie et al. 2010), composition maps based on visible HST maps and visible and near-IR spectra (Grundy & Fink 1996; Grundy & Buie 2001), and rotationally resolved thermal emission (Lellouch et al., 2000, 2011).

NASA's New Horizons spacecraft will fly by the Pluto system in July of 2015 (Stern et al., 2008). Much of the planning is based on the expectation, from Hansen & Paige (1996) models and occultation observations, that the atmosphere now through encounter is in a slowly changing pressure plateau. However, models and computers from the mid-1990's limited the number of cases that could be investigated by Hansen & Paige (1996). If the pressure plateau ends near or before 2015, this will have profound implications for the world that New Horizons will encounter in 2015, and our ability to relate this snapshot to preceding or following observations. For this reason, we have developed new volatile transport models with application to Pluto (Young, 2012; Young, in prep), and compared them to the existing occultation record.

## 2. Volatile Transport Model

This study uses the three-dimensional volatile-transport (VT3D) code developed in Young (2012) and Young (in prep). Energy balance is identical to that used by Hansen & Paige (1996) and Young (2012). Energy is balanced locally between (i) insolation, (ii) thermal emission, (iii) conduction, (iv) internal heat flux, and, in areas covered by solid $N_2$, (v) latent heat of sublimation and (vi) specific heat needed to raise the temperature of the volatile slab. The internal heat flux is taken to be 6 erg cm$^{-2}$ s$^{-1}$, following Hansen & Paige (1996). The latent heat of crystallization of the $N_2$ phase change at 32.6 K has a minor effect on the seasonal variation of Pluto or Triton (Spencer & Moore, 1992) and is ignored in this paper.

For current-day Pluto, and for much of Pluto's orbit, Pluto's atmosphere effectively transports both mass and energy (in the form of latent heat) from areas of high to low insolation (Stern et al., 1988, Spencer et al., 1997). In this case, the volatile ice temperature is nearly uniform over the entire body, as is the surface pressure. Conservation of mass, integrated over the entire body, is used to eliminate the latent-heat terms in the energy equations (Young 2012). When Pluto's atmosphere is too tenuous to maintain an isothermal, isobaric surface, VT3D treats the surface as a splice between areas with efficient transport, which share a common volatile ice temperatures and surface pressure, and areas with no lateral transport of volatiles, where ice temperatures follow strictly local energy balance.

Pluto's ice temperature should vary only minimally over a Pluto day (Young 2012), so this paper averages solar insolation over latitude bands. Simulations were initialized at aphelion, with the specified $N_2$ inventory distributed evenly over the surface. Surface and subsurface temperatures were initialized using a sinusoidal decomposition of solar forcing, as described in Young (2012) and Young (in prep).

Temperatures within the substrate were calculated at 2.5 points per skin depth, down to 7.2 skin depths. Temperatures were calculated on a relatively short time grid of 240 per Pluto year, or just over 12 Earth months per timestep. With this fine a time step, the explicit forward-timestep is stable, and was used in the calculations presented here. Because of the improved initial conditions, only three Pluto years were needed before the simulations converged (that is, the $N_2$ ice temperatures in the second and third years differed by only a few percent of the peak-to-peak seasonal variation).

## 3. Parameter Space Search

Calculation of a single Pluto simulation in the above manner is very fast, allowing a wide parameter space search. The bolometric hemispheric albedo, $A_V$, of the $N_2$ ice was varied from 0.2 to 0.8 in steps of 0.2, for 7 values. This range matches the range of values used by Hansen & Paige (1996), and includes the values described as good or acceptable fits by Lellouch et al. (2011). The emissivity of the $N_2$ ice, $\varepsilon_V$, was calculated at only two values, 0.55 and 0.8. The lower value is the value adopted by Young (2012), based on Lellouch et al. (2011), while 0.8 emissivity is the highest considered by Hansen & Paige (1996). The substrate bolometric hemispheric albedo, $A_S$, was fixed at 0.2 for all runs, based on the rough agreement of runs 12, 34 and 38 of Hansen & Paige (1996) with the occultation record; results are not sensitive to changes in the substrate albedo, as long as it is low. All runs used a substrate bolometric emissivity, $\varepsilon_S$, of 1.0, based on the "good fits" of Lellouch et al. (2011). The thermal inertia, $\Gamma$, was varied logarithmically at 9 values between 1 and $10^4$ J m$^{-2}$ s$^{-1/2}$ K$^{-1}$. (MKS units are used for $\Gamma$ for convenience and comparison with recent literature). This range is a superset of the values modeled by Hansen & Paige (1996) (41 to $2.1 \times 10^3$ J m$^{-2}$ s$^{-1/2}$ K$^{-1}$), and includes the thermal inertia derived by Lellouch et al. (2011) (~18 J m$^{-2}$ s$^{-1/2}$ K$^{-1}$). Six values of the total $N_2$ inventory, $m_{N2}$, ranged from 2 to 64 g cm$^{-2}$, varying by factors of 2, a range that includes the values modeled by Hansen & Paige (1996).

All simulations were passed through a wide sieve to identify those results roughly consistent with stellar occultations in 1988 and 2006. More detailed comparisons are given in Section 4. The range of acceptable pressures for 2006 was taken to be 7 to 78 µbar. The lower end of the range is dictated by the fact that occultations in 2006 probed down to at least 6 µbar (Young et al., 2008). The upper end of the range is guided by Lellouch et al., (2009), who combined high-resolution IR spectra of Pluto's gaseous $CH_4$ with stellar occultations to derive a maximum pressures in 2008 of 24 µbar. The larger upper end of the 2006 sieve range accounts for the difference in time between 2006 and 2008, and the model dependence of the Lellouch et al. (2009) result.

Young et al. (2008) report that the pressure in 2006 at a reference radius of 1275 km from Pluto's center was a factor of 2.4 ± 0.3 times larger than in 1988. Taking into account the difficulty in relating pressure at 1275 km to Pluto's surface, spanning a gap of some 75 to 100 km, the sieve requires a ratio of the 2006 and 1988 surface pressures in the range of 1.5 to 3.1. The limits on the ratio of pressures would imply a range for 1988 of 2.2 to 52 µbar. However, the stellar occultation of 1988 provides an additional constraint, as it probed to 3.0 µbar. The final 1988 pressure range for the sieve is 3.0 to 52 µbar.

Of the 756 simulations, 53 (7%) matched the coarse sieve (Table 1). The volatile migration patterns were visually inspected for each of these 53 runs, and were found to fall into one of three categories. The first category, called *permanent northern volatiles* (PNV), (Fig. 1, top), with 26 runs, had volatiles on the northern (current summer) hemisphere throughout the entire Pluto year. In general, these runs have large thermal inertia, and often have large volatile inventories (Table 1; Fig 2, left). Fig 1 (top) plots a typical example. All of the runs that have volatiles on the northern hemisphere at all times in the season have gradual pressure changes; about half have pressures between 10 and 100 µbar throughout the entire year, and all have minimum pressures above 0.4 µbar.

The other half of the simulations that matched the coarse sieve have complete exchanges of volatiles between the northern and southern hemispheres, with each hemisphere becoming completely bare at some time of Pluto's season (Fig. 1, middle and bottom). These generally have larger variations in pressure than the PNV cases, usually with two distinct pressure maxima, one near the southern summer solstice, and one between perihelion equinox and northern summer solstice. These break into two subcases. Some of these have two volatile caps for a long period after the perihelion equinox, defining the second category, *exchange with pressure plateau* (EPP). Others lose the northern volatiles shortly after perihelion, defining the third category, *exchange with early collapse* (EEC). Both the EPP and EEC cases generally have moderate to small values of thermal inertia, with the EEC cases having smaller volatile inventories than the EPP cases (Table 1; Fig 2, middle and right).

As with the PNV runs, the southern summer hemisphere gets quickly denuded near southern summer solstice, giving a period of cooling northern winter volatiles. Also as with the PNV runs, there is a period of rising pressures before perihelion equinox as the northern hemisphere gets more direct illumination. At this point, the cases with exchange of volatiles deviates from the PNV cases. For the exchange cases, a southern volatile cap forms near the perihelion equinox. A period of exchange between the northern summer cap and the new southern winter cap ensues, with relatively stable surface pressures. The post-perihelion volatile migration in the exchange cases mirrors the post-aphelion migration: the summer (northern) hemisphere disappears, the winter (southern) cap cools, the southern cap becomes more directly illuminated (transitioning from winter to summer), followed finally near aphelion by the exchange to a new winter (northern) hemisphere. The distinction between an EPP or an EEC case is based on the state of the northern volatile cap at the time of the New Horizons encounter in mid-2015. A typical EPP run is shown in the middle panel of Fig. 1, and a typical EEC is shown in the lower panel.

## 4. Predictions for New Horizon

Plotting the predicted surface pressure on decadal timescales (Fig 3, left), the PNV cases show a general trend of gradual rising near perihelion, perhaps followed by an equally gradual decrease. The line with the largest decrease between 2010 and 2020 is run PNV11, which has the smallest thermal inertia (100 J m$^{-2}$ s$^{-1/2}$ K$^{-1}$) of all the runs that had northern volatiles throughout the Pluto year. Also plotted in Fig. 3 are pressures derived from occultations since 1988 (Table 2). Pressures are reported here at a reference radius of 1275 km from Pluto's center, since stellar occultations do not probe to Pluto's surface. Pressures at the reference radius are plotted as open circles for pressures derived from simple model fits to the data (e.g., Young et al., 2008), or with open diamonds for pressures derived from physical models (Zalucha et al., 2011). In the physical models, surface pressures (open squares, from Zalucha et al., 2011) are larger than the reference pressures by a factor of 5.5 to 10. Estimates of the surface pressures from simple model fits are plotted by multiplying their corresponding reference pressures by a factor of 7.2 (solid black circles) or 32 (solid gray circles); the factors were chosen to scale within the sieve for 1988. Zalucha et al. (2011) show that physical models with surface pressures that differ by a factor of three can give essentially the same half-light shadow radius, a value often used as a proxy for pressure at the reference level. Therefore, we adopt estimated systematic errors of sqrt(3) in the surface pressure from simple models. Clearly, the PNV pressures are in general agreement with the occultation record. The values of the thermal inertia for the PNV cases are high (Fig 2), with nearly all in the range 316 - 3160 J m$^{-2}$ s$^{-1/2}$ K$^{-1}$. This can be compared with the thermal inertia for pure, solid, H$_2$O of ~2100 J m$^{-2}$ s$^{-1/2}$ K$^{-1}$ (Spencer & Moore, 1992). The values of $\Gamma$ for PNV runs are high, but not implausible, especially considering that the skin depth for these larger thermal inertia values is 100 m or more (Hansen & Paige, 1996). Given that Lellouch et al. (2011) find $\Gamma$ ~18 J m$^{-2}$ s$^{-1/2}$ K$^{-1}$ for the diurnal wave, a PNV solution would require an increase of $\Gamma$ with depth.

All PNV solutions predict surface pressures greater than 10 µbar in 2015. Since the Alice and REX instruments on New Horizons, and the planned observations at encounter, were designed for surface pressures of 4 µbar or more, these pressures are well above the design specifications for the Alice and REX measurements. Most of the PNV solutions have no volatiles on the southern hemisphere near or shortly after the perihelion equinox. The implication is that, for the decades before the New Horizons encounter, much of the volatile migration will be from the directly illuminated high northern latitudes to the less directly illuminated edges of the northern volatile cap. The result may well be similar to that which Voyager saw at Triton, showing an old cap with a collar of new frost.

The surface pressures from the EPP category are also consistent with the occultation record, within the measurement and modeling uncertainties of the occultations. With the exception of the one high-$\Gamma$ EPP1, these all have thermal inertia less than or equal to 10 J m$^{-2}$ s$^{-1/2}$ K$^{-1}$. This is lower than the diurnal value of ~18 J m$^{-2}$ s$^{-1/2}$ K$^{-1}$ measured by Lellouch et al. (2011). As it is unlikely that $\Gamma$ decreases with depth, it is likely that only EPP1, EPP2, EPP6, and EPP12 are plausible solutions in this catagory. For much of Pluto's orbit, especially near perihelion, the volatile migration pattern of EPP1 is similar to the other high-inertia PNV cases. EPP1 is represented in Fig 3, middle, by the line with the gradual pressure changes. For other EPP cases, New Horizons might see an old, summer, northern pole, with just a sliver of the new, southern, winter pole at latitudes poleward of $-15°$. The predicted surface pressure for EPP2, EPP6, and EPP12 is 15-25 µbar, also well above the design specifications for REX and Alice instruments.

The runs in the EEC category are only consistent with the occultation record because of the difficultly in relating pressures at occultation altitudes to surface pressures.

Application of physical atmospheric models, such as those by Zalucha et al., (2011), to occultation observations may decrease the errors and allow this case to be eliminated. Because the atmospheric pressure decreases rapidly in EEC cases, observations in 2011–2015 will be particularly diagnostic. An occultation in hand from 2011 June 23 had chords all one side of the occultation midline, making geometric reconstruction too inaccurate for use here. Another occultation from 2012 Sep 9 is currently being analyzed. Most of the EEC runs have reasonable values of $\Gamma$.

The runs in the EEC category all predict surface pressures less than 1 μbar. Despite the low pressures, only one case, EEC12, has surface pressures too small to support global atmosphere. This is because volatile migration is only from the edge of the winter cap toward the winter pole. The mass and the distances are small, so winds are subsonic even for EEC9, EEC10, and EEC11. For all the EEC cases, essentially by definition, the northern, summer volatile cap is completely or nearly completely sublimated. In most cases, the southern, winter volatile cap only extends to roughly –30°. There will be few $N_2$-rich volatiles to be observed by the LEISA instrument on New Horizons. Note, however, that this version of the model does not track the $CH_4$-rich volatiles, and these may remain on the visible hemisphere. The Alice measurement of $N_2$ opacity is effective even at these lower pressures, but, if the EEC models are correct, the REX instrument will measure near-surface pressures and temperatures with degraded sensitivity.

## 5. Future Work

Hansen (personal communication) has recently rerun the models of Hansen & Paige (1996). Vangvichith & Forget (2011) have presented simulations for Triton that include seasonal volatile migration in global circulation models. It would be most instructive to compare of results of different volatile transport models with identical physical parameters.

Pluto has an albedo and lightcurve record extending back to 1933 (Scheafer et al., 2008), an almost continuous record of near-IR spectroscopy since 1995 (Grundy & Buie 2001), and rotationally resolved thermal observations 1997-2010 (Lellouch et al., 2011). Comparisons of volatile transport models with these observations can further constrain model assumptions. In most cases, these comparisons will require us to treat other volatiles in addition to $N_2$.

Atmospheric models, such as Zalucha et al., (2011) may prove the key to relating pressures at occultation altitudes to pressures at the surface.

Continuing ground-based observations of Pluto's albedo, spectra, and atmosphere will provide a temporal context in which to place the New Horizons flyby data. Conversely, New Horizons will provide a rich data set with which to understand Pluto's seasonal evolution, including visible maps (which may show a Triton-like collar for PVN, an old summer and new winter pole for EPP, and a lack of sunlit $N_2$ ice for EEC); composition maps (directly revealing the location of the volatiles); Pluto's radius (needed for interpreting the stellar occultation record); and the atmospheric temperature structure.

This work was supported in part by NASA's New Horizons mission to the Pluto system.

- 14 -

Fig. 1. Top: a typical run for the Permanent Northern Volatiles (PNV) category, run PNV23. Middle: a typical run for the Exchange with Pressure Plateau (EPP) category, run EPP6. Bottom: a typical run for the Exchange with Early Collapse (EEC) category, run EEC6. For each category, the plot on the left shows Pluto over a season. Pluto's orbit is show to scale, with time segments of one-twelfth of an orbit marked in alternating shades of gray. The circles represent Pluto at each of 12 times in the orbit, indicated by date starting at the previous aphelion. The short vertical bar behind the circles represents the rotational axis, oriented so that the axis is perpendicular to the sun vector at the equinoxes, with the northern pole at the top (currently pointed sunward). Latitude bands are colored with their geometric albedos. The plots on the left show geometric albedo and surface pressure, $p$, as a function of year. Note the change in pressure scale for the top plot.

Fig. 2. Parameters for PNV, EPP, and EEC categories. Circles are centered on the corresponding hemispheric albedo ($A_V$) and thermal inertia ($\Gamma$). Circle sizes relate to the total $N_2$ inventory, ranging form 2 g cm$^{-2}$ (smallest circles) to 64 g cm$^{-2}$ (largest circles).

Fig 3. Predicted surface pressures for the PNV (left), EPP (middle), and EEC (right) cases. Open circles: occultations pressures at 1275 km derived by simple model fitting with 1-$\sigma$ error bars (Table 2). Filled circles: estimated surface pressures, scaling the reference pressures by factors of 7.2 (black) and 31 (gray), with plotted error bars indicating a systematic contribution to the error in the surface pressure of 1.7 (see text). Diamonds: pressures at 1275 km derived from fitting physical atmospheric models to occultations (Zalucha et al., 2011). Squares: two values of the surface pressures derived from fitting physical atmospheric models to occultations, under the "troposphere excluded" (lower square) and "troposphere included" assumptions (Zalucha et al., 2011).

Table 1. Runs that pass the coarse sieve, sorted by 2015 pressure within each category.

| Run | $A_V$ | $\varepsilon_V$ | $\Gamma$ J m$^{-2}$ s$^{-1/2}$ K$^{-1}$ | $m_{N2}$ g cm$^{-2}$ | Surface Pressure, μbar | | | |
|---|---|---|---|---|---|---|---|---|
| | | | | | 1988 | 2002 | 2006 | 2015 |
| PNV1  | 0.50 | 0.80 | 3160.  | 16 | 36  | 63  | 75  | 102 |
| PNV2  | 0.60 | 0.55 | 3160.  | 2  | 35  | 58  | 68  | 95  |
| PNV3  | 0.50 | 0.80 | 10000. | 8  | 50  | 60  | 64  | 73  |
| PNV4  | 0.50 | 0.80 | 3160.  | 4  | 22  | 40  | 49  | 73  |
| PNV5  | 0.50 | 0.80 | 3160.  | 2  | 19  | 36  | 45  | 69  |
| PNV6  | 0.60 | 0.80 | 1000.  | 64 | 26  | 54  | 60  | 59  |
| PNV7  | 0.50 | 0.80 | 10000. | 2  | 39  | 47  | 51  | 58  |
| PNV8  | 0.70 | 0.55 | 3160.  | 16 | 29  | 40  | 44  | 53  |
| PNV9  | 0.70 | 0.55 | 3160.  | 32 | 33  | 44  | 47  | 52  |
| PNV10 | 0.70 | 0.55 | 3160.  | 8  | 26  | 37  | 42  | 52  |
| PNV11 | 0.70 | 0.80 | 100.   | 32 | 32  | 60  | 55  | 36  |
| PNV12 | 0.60 | 0.80 | 3160.  | 16 | 14  | 22  | 26  | 34  |
| PNV13 | 0.60 | 0.80 | 3160.  | 32 | 16  | 25  | 28  | 34  |
| PNV14 | 0.60 | 0.80 | 3160.  | 8  | 12  | 20  | 24  | 33  |
| PNV15 | 0.70 | 0.55 | 3160.  | 2  | 13  | 20  | 24  | 32  |
| PNV16 | 0.60 | 0.80 | 3160.  | 64 | 20  | 27  | 29  | 32  |
| PNV17 | 0.70 | 0.55 | 3160.  | 4  | 13  | 19  | 22  | 29  |
| PNV18 | 0.60 | 0.80 | 3160.  | 4  | 7.5 | 13  | 15  | 22  |
| PNV19 | 0.70 | 0.80 | 316.   | 64 | 15  | 28  | 27  | 21  |
| PNV20 | 0.80 | 0.55 | 316.   | 32 | 9.9 | 21  | 22  | 19  |
| PNV21 | 0.60 | 0.80 | 3160.  | 2  | 7.1 | 12  | 14  | 18  |
| PNV22 | 0.80 | 0.55 | 316.   | 64 | 13  | 18  | 17  | 15  |
| PNV23 | 0.70 | 0.80 | 1000.  | 32 | 3.9 | 9.2 | 11  | 15  |
| PNV24 | 0.70 | 0.80 | 1000.  | 64 | 6.0 | 10  | 11  | 12  |
| PNV25 | 0.80 | 0.55 | 1000.  | 16 | 3.5 | 6.9 | 8.2 | 11  |
| PNV26 | 0.80 | 0.55 | 1000.  | 32 | 5.0 | 8.0 | 8.8 | 10  |

Table 1, cont. Runs that pass the coarse sieve, sorted by 2015 pressure within each category.

| Run | $A_V$ | $\varepsilon_V$ | $\Gamma$ J m$^{-2}$ s$^{-1/2}$ K$^{-1}$ | $m_{N2}$ g cm$^{-2}$ | Surface Pressure, μbar | | | |
|---|---|---|---|---|---|---|---|---|
| | | | | | 1988 | 2002 | 2006 | 2015 |
| EPP1  | 0.50 | 0.80 | 3160. | 8  | 30  | 55  | 67 | 98 |
| EPP2  | 0.80 | 0.55 | 10.   | 16 | 17  | 50  | 44 | 25 |
| EPP3  | 0.70 | 0.80 | 3.    | 16 | 19  | 51  | 45 | 25 |
| EPP4  | 0.80 | 0.55 | 3.    | 16 | 19  | 52  | 45 | 25 |
| EPP5  | 0.70 | 0.80 | 1.    | 16 | 20  | 49  | 45 | 25 |
| EPP6  | 0.70 | 0.80 | 10.   | 16 | 16  | 49  | 44 | 25 |
| EPP7  | 0.80 | 0.55 | 1.    | 16 | 19  | 53  | 45 | 25 |
| EPP8  | 0.80 | 0.55 | 3.    | 8  | 13  | 40  | 35 | 19 |
| EPP9  | 0.80 | 0.55 | 1.    | 8  | 14  | 41  | 33 | 19 |
| EPP10 | 0.60 | 0.80 | 3.    | 8  | 42  | 106 | 77 | 18 |
| EPP11 | 0.60 | 0.80 | 1.    | 8  | 45  | 109 | 77 | 17 |
| EPP12 | 0.70 | 0.80 | 10.   | 8  | 9.9 | 35  | 31 | 15 |
| EPP13 | 0.70 | 0.80 | 3.    | 8  | 12  | 34  | 31 | 13 |
| EPP14 | 0.80 | 0.55 | 3.    | 4  | 7.6 | 26  | 19 | 4.6 |
| EPP15 | 0.80 | 0.55 | 1.    | 4  | 8.7 | 25  | 19 | 4.5 |
| EEC1  | 0.70 | 0.55 | 10.   | 4  | 31  | 97  | 74 | 0.98 |
| EEC2  | 0.70 | 0.80 | 10.   | 4  | 4.8 | 19  | 12 | 0.28 |
| EEC3  | 0.70 | 0.55 | 3.    | 4  | 39  | 97  | 71 | 0.28 |
| EEC4  | 0.70 | 0.55 | 1.    | 4  | 41  | 97  | 69 | 0.21 |
| EEC5  | 0.20 | 0.80 | 100.  | 2  | 13  | 147 | 28 | 0.18 |
| EEC6  | 0.70 | 0.80 | 3.    | 4  | 6.5 | 17  | 12 | 0.078 |
| EEC7  | 0.60 | 0.80 | 32.   | 4  | 9.3 | 48  | 30 | 0.053 |
| EEC8  | 0.70 | 0.80 | 1.    | 4  | 6.9 | 18  | 13 | 0.029 |
| EEC9  | 0.50 | 0.55 | 32.   | 2  | 51  | 207 | 70 | 0.019 |
| EEC10 | 0.60 | 0.55 | 32.   | 2  | 20  | 103 | 44 | 0.015 |
| EEC11 | 0.50 | 0.80 | 32.   | 4  | 24  | 100 | 57 | 0.014 |
| EEC12 | 0.60 | 0.80 | 10.   | 4  | 19  | 47  | 29 | 7.6E-04 |

Table 2. Pressures at reference altitude 1275 km from Pluto's center, measured by stellar occultation.

| Date | Pressure at 1275 km, μbar | Reference |
| --- | --- | --- |
| 1988 Jun 9 | $0.83 \pm 0.11$ | Elliot & Young 1992 |
|  | $1.4^{+0.03}_{-0.05}$ | Zalucha et al., 2011 |
| 2002 Aug 21 | $1.76 \pm 0.51$ | Elliot et al., 2003 |
|  | $1.8^{+1.7}_{-0.7}$ | Zalucha et al., 2011 |
| 2006 Jun 12 | $1.86 \pm 0.10$ | Young et al., 2008 |
|  | $2.4^{+0.08}_{-0.07}$ | Zalucha et al., 2011 |
| 2007 Mar 18 | $2.03 \pm 0.2$ | Person et al., 2008 |
| 2007 Jul 31 | $2.09 \pm 0.09$ | Olkin et al., in prep |
| 2008 Aug 25 | $4.11 \pm 0.54$ | Buie et al., 2009 |
| 2009 Apr 21 | $2.59 \pm 0.09$ | Young et al., 2009 |
| 2010 Feb 14 | $1.787 \pm 0.076$ | Young et al., 2010 |

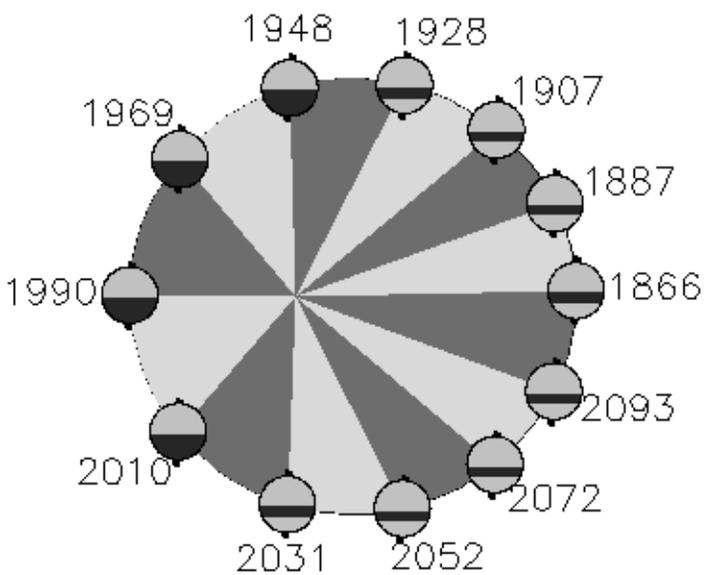
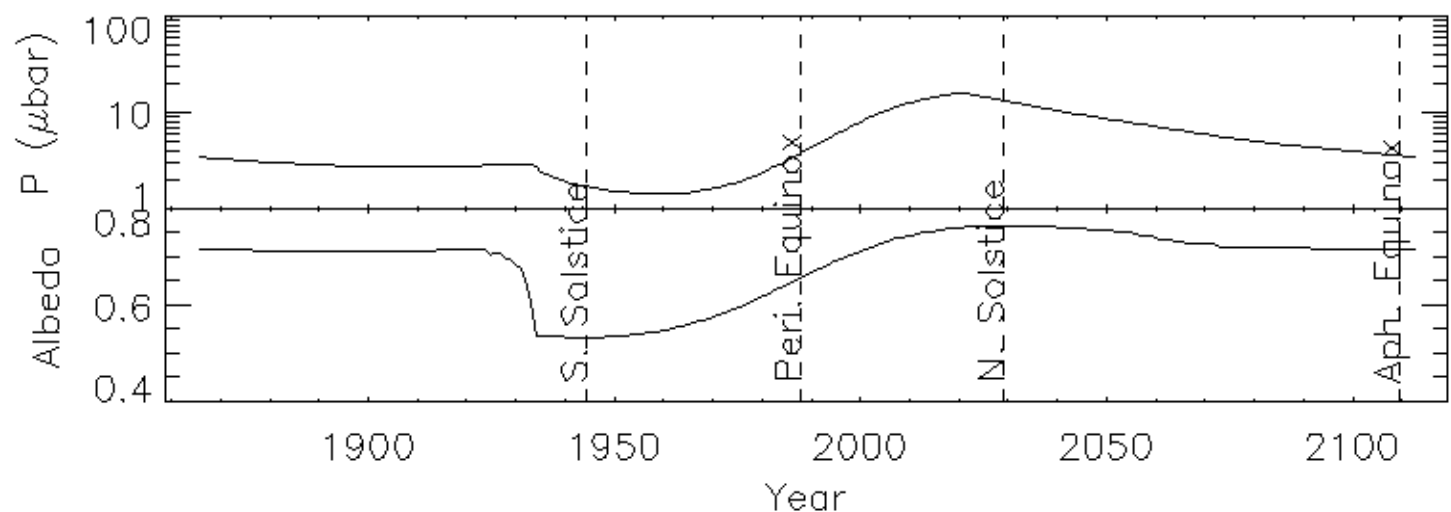
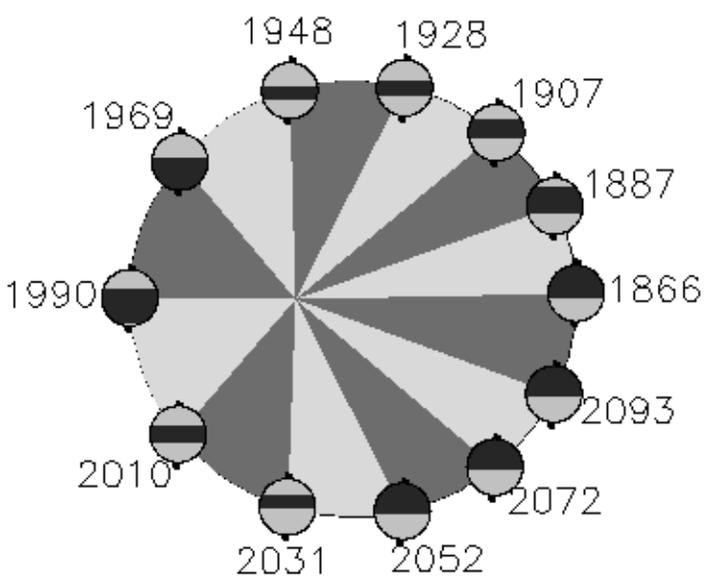
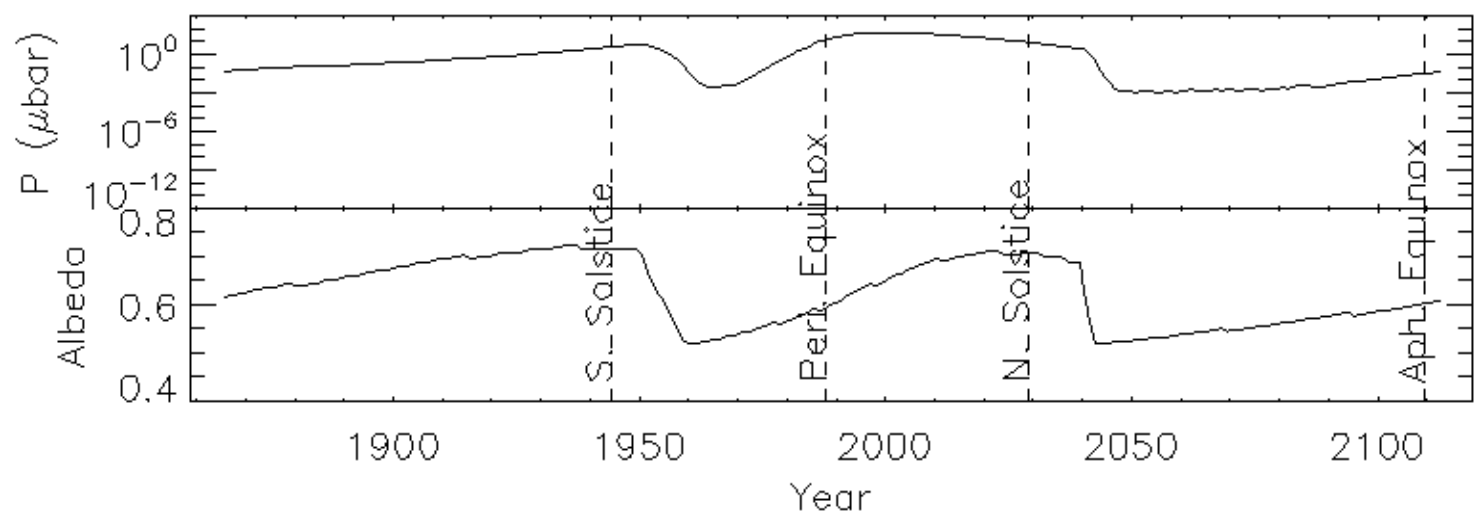
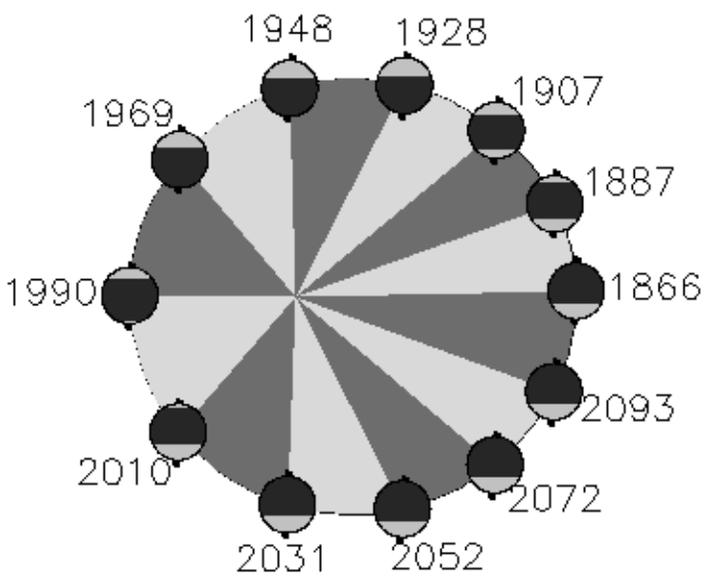
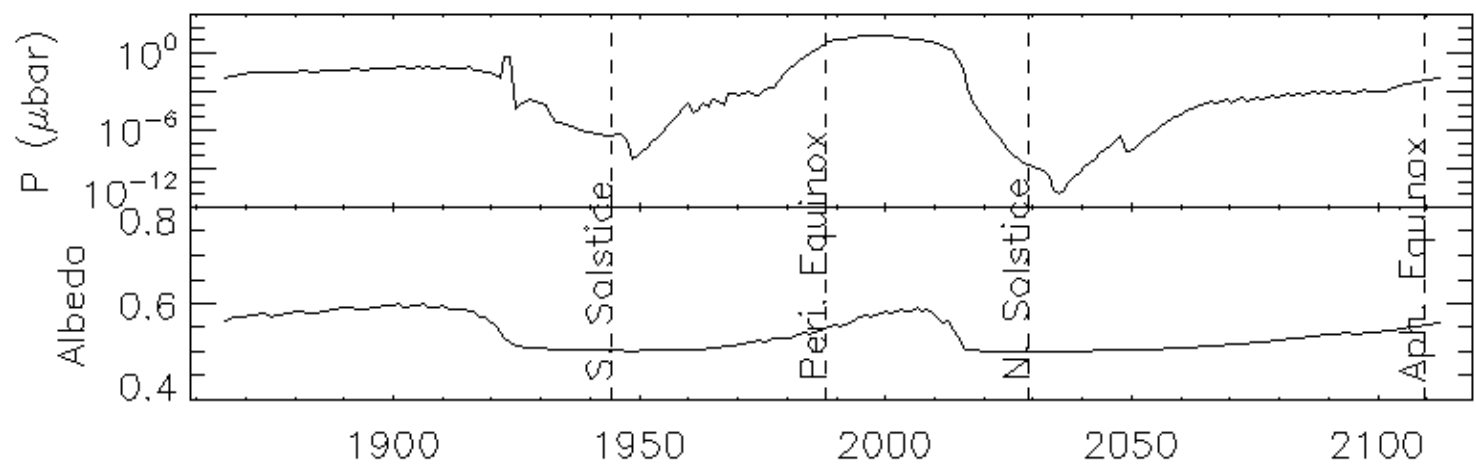

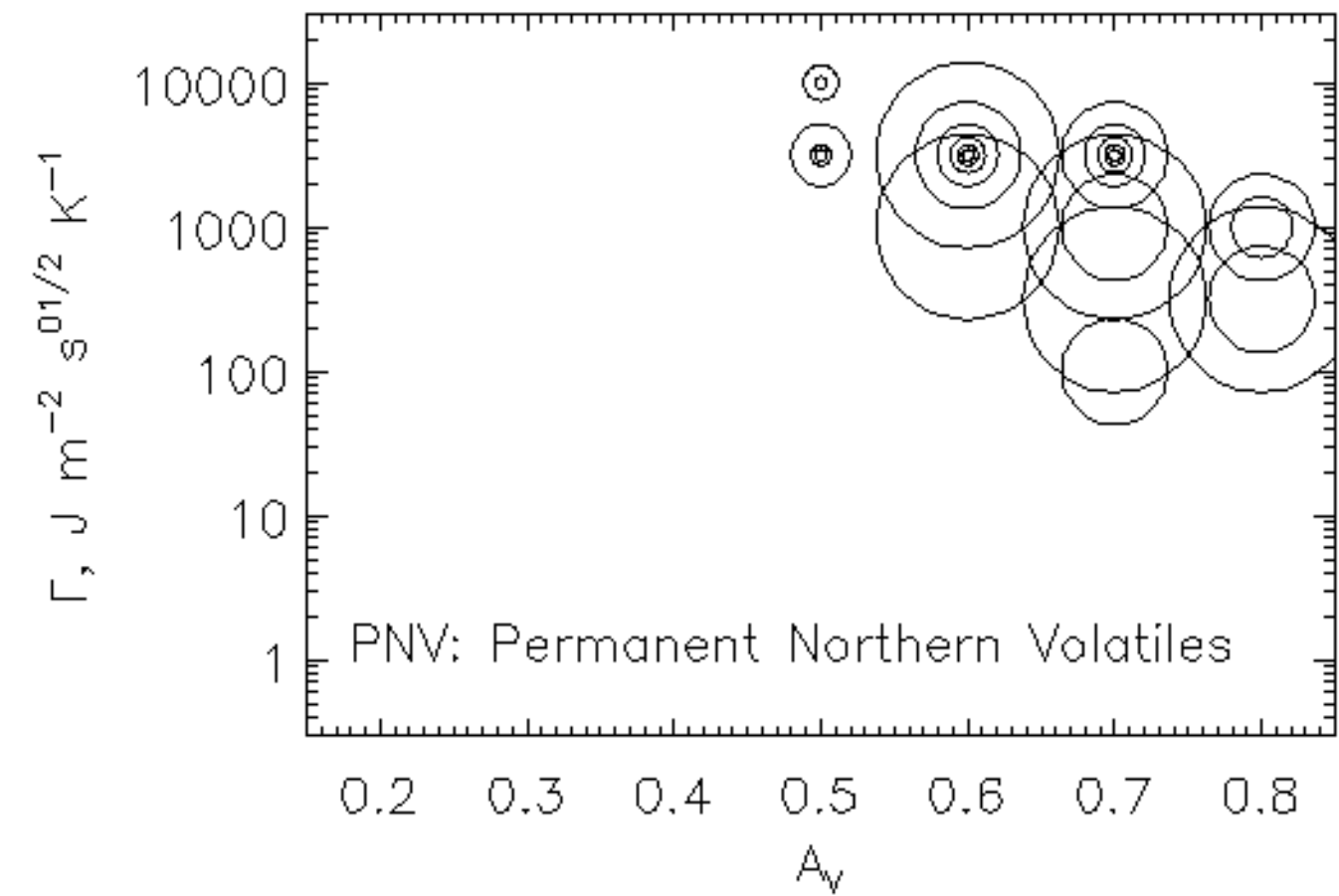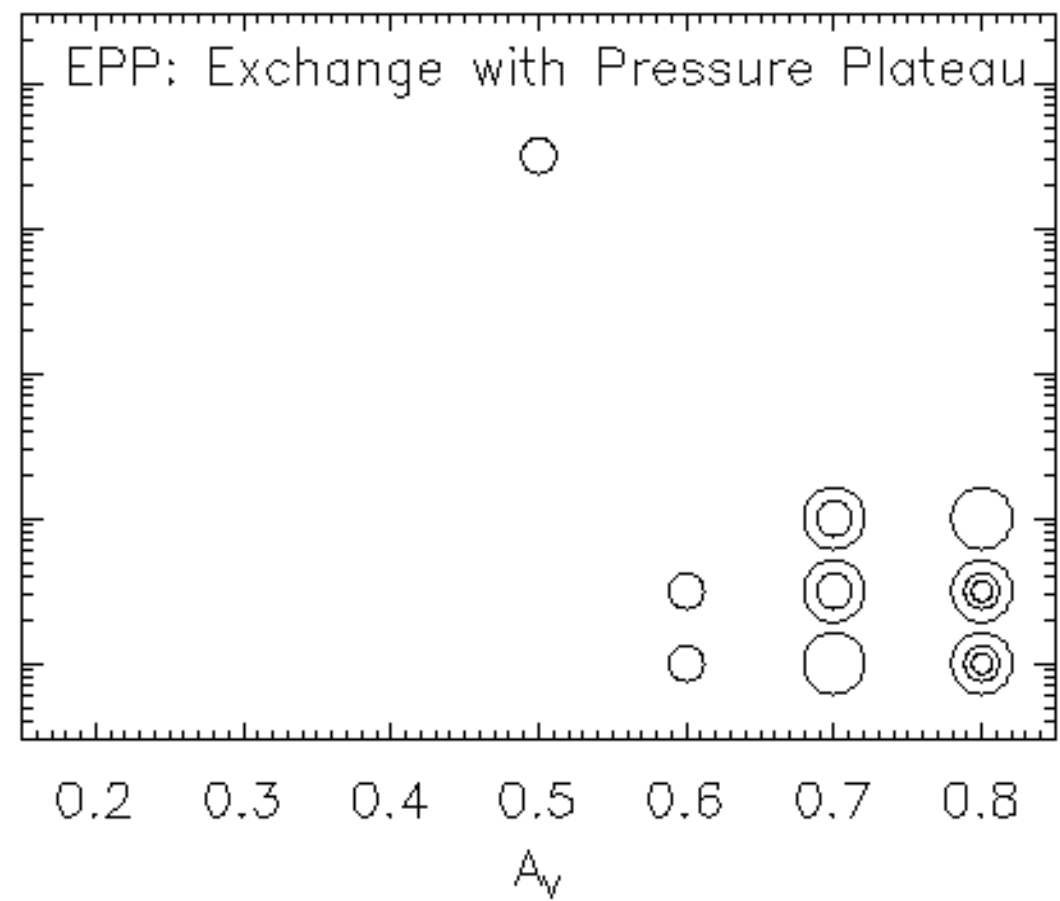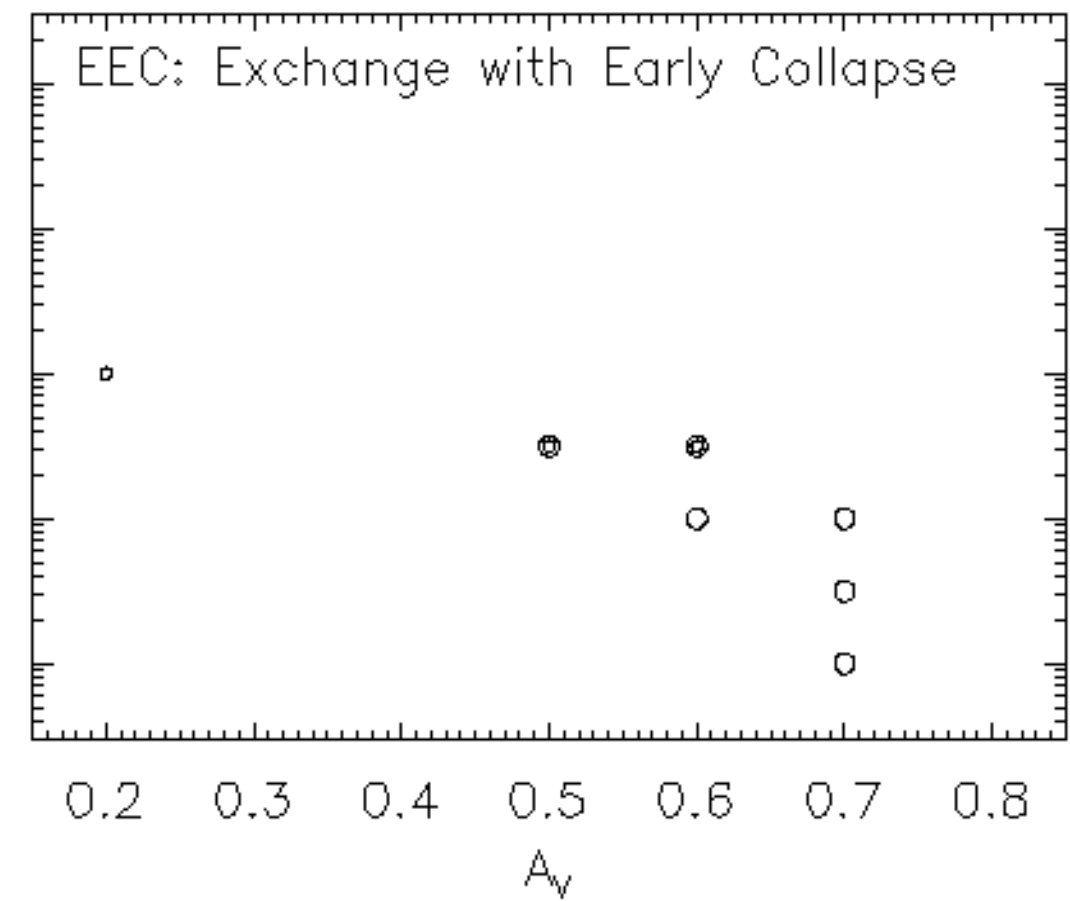

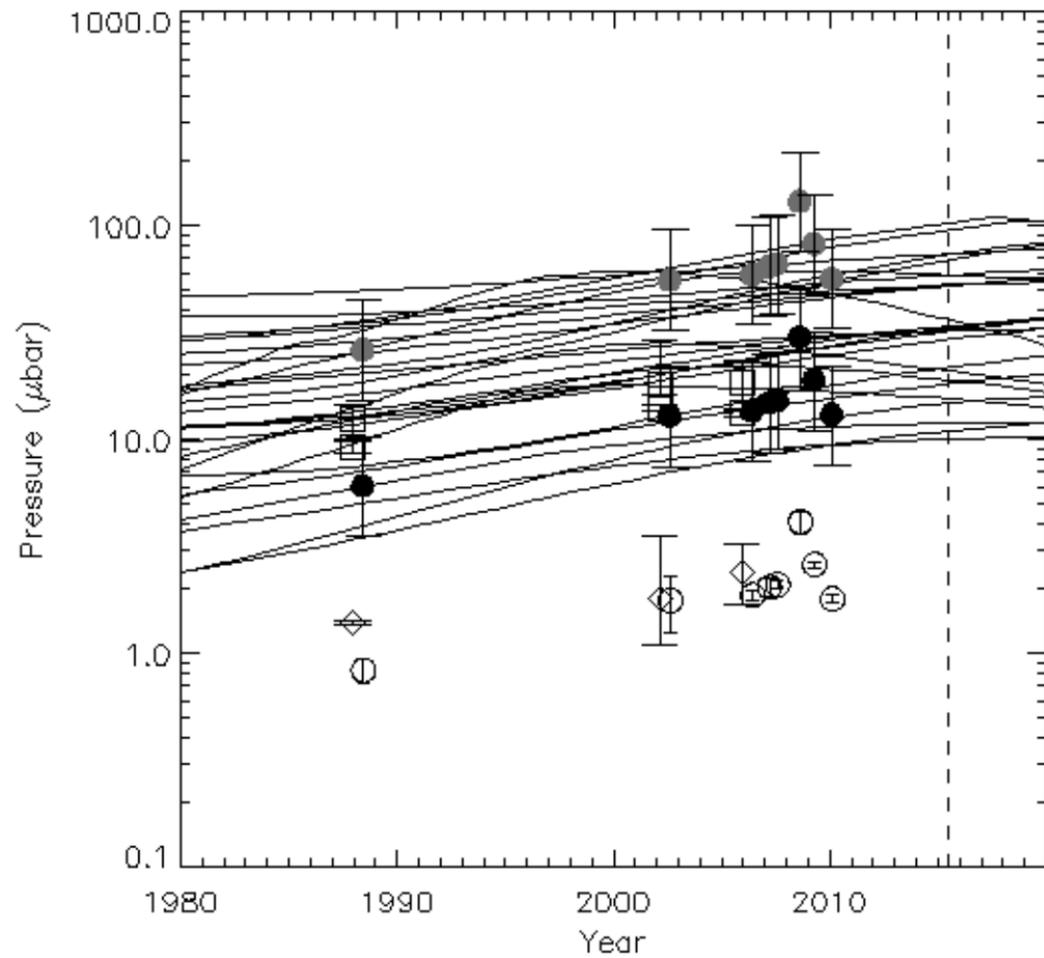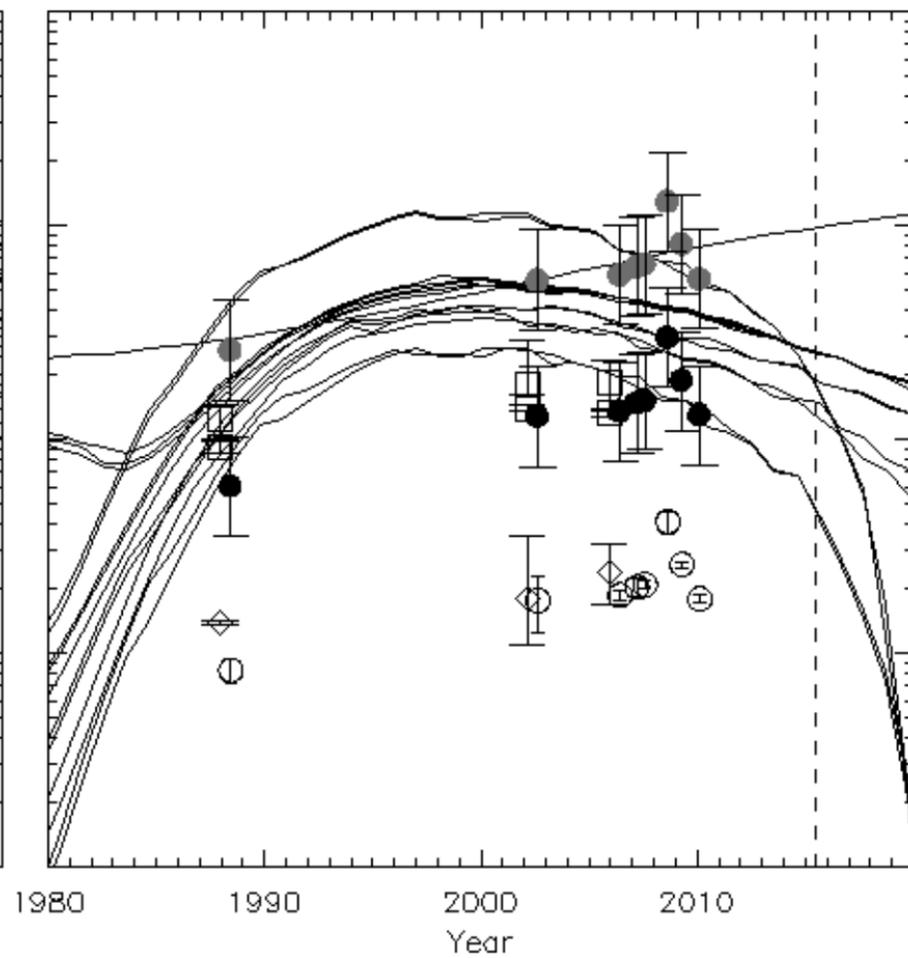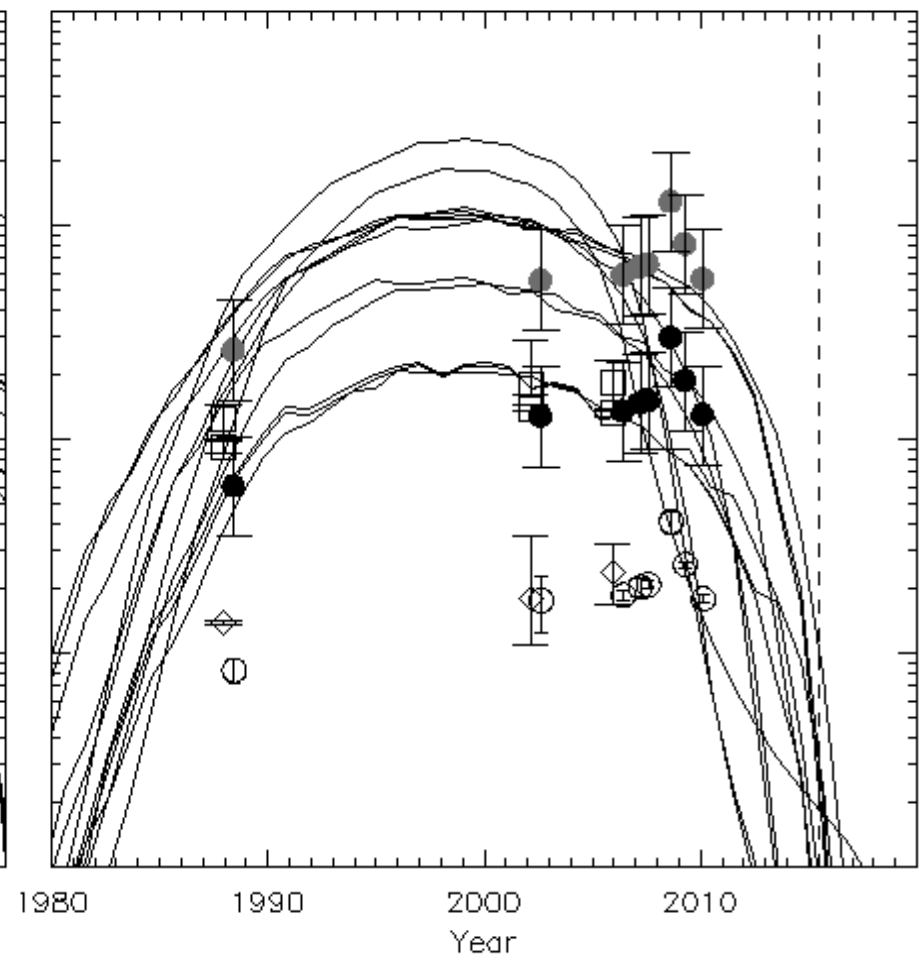